\def\th{\theta}

\def\vf{\varphi}
\def\cA{${\cal A}$}
\def\cB{${\cal B}$}
\def\cC{${\cal C}$}
\def\hbar{{\mathchar'26\mkern-9mu h}}
\def\a{\alpha}
\def\b{\beta}

\def\d{\delta}

\def\g{\gamma}
\def\l{\lambda}
\def\Re{{\rm Re}}

\documentstyle[12pt]{article}
\advance\textwidth 1in
\advance\oddsidemargin -0.5in
\advance\textheight 1in
\advance\topmargin -0.5in
\title{Convergence of the Optimized Delta Expansion for the Connected
Vacuum Amplitude: Zero Dimensions}

\author{\\ \\Carl M. Bender\\
        Physics Department\\
        Washington University\\
        St. Louis, MO \ \ 63130 USA \\
        \\
     Anthony Duncan\\
     Department of Physics and Astronomy \\
     University of Pittsburgh \\
     Pittsburgh, PA \ \ 15260 USA \\
     \\
     H.F. Jones \\
     Physics Department\\
     Imperial College \\
     London, SW7 2BZ \ \ UK}
\begin{document}
\maketitle
\def\thepage{Imperial/TP/92-93/55}
\thispagestyle{myheadings}
\newpage
\baselineskip=24pt
\pagenumbering{arabic}
\addtocounter{page}{1}
\begin{abstract}
Recent proofs of the convergence of the linear delta expansion in zero and in
one dimensions have been limited to the analogue of the vacuum generating
functional in field theory. In zero dimensions it was shown that with an
appropriate, $N$-dependent, choice of an optimizing parameter $\l$,
 which is an important
feature of the method, the sequence of approximants $Z_N$ tends to $Z$ with an
error proportional to ${\rm e}^{-cN}$. In the present paper we establish the
convergence of the linear delta expansion for the connected vacuum function
$W=\ln Z$. We show that with the same choice of $\l$ the corresponding sequence
$W_N$ tends to $W$ with an error proportional to ${\rm e}^{-c\sqrt N}$. The
rate
of convergence of the latter sequence is governed by the positions of the zeros
of $Z_N$.
\bigskip

\end{abstract}
\newpage
\baselineskip=24pt
\subsection*{1. Introduction}
\label{sec:introduction}

In a previous paper \cite{LDEZ0} it was proved that in zero-dimensional
space-time the optimized linear delta expansion can completely cure the
problems
inherent in conventional perturbation theory. A conventional perturbation
expansion is a formal series in powers of the coupling constant. Such a power
series typically has a zero radius of convergence and sometimes is not even
useful as an asymptotic series because it is not Borel summable. In contrast,
the optimized linear delta expansion produces a
sequence of approximants which converge
rapidly to the exact answer, with an error $R_N$ that decreases exponentially
with the order $N$: $R_N \sim \exp(-cN)$. The proof given in Ref.~\cite{LDEZ0}
was subsequently extended \cite{LDEZ1} to one-dimensional space-time (the
quantum anharmonic oscillator), where it was shown that for the
finite-temperature partition function $R_N \sim \exp(-cN^{2/3})$. The
technique can also be extended to establish a proof of convergence for
cutoff $\vf^{4}_{2,3}$ theories (with either sign of the squared mass)
in finite volume.

However, the proofs of convergence given in Refs.~\cite{LDEZ0} and \cite{LDEZ1}
were limited to the partition function (vacuum-vacuum amplitude) $Z$, which
represents the sum of all vacuum graphs, disconnected as well as connected. In
quantum field theory the crucial quantity to compute is $W$, the logarithm of
the vacuum-vacuum function, which represents the sum of the connected vacuum
graphs only. It is nontrivial to show that the optimized linear delta expansion
converges for $W$ because $Z_N$, the $N$th approximation to $Z$, has zeros in
the complex-$\d$ plane at which $\ln Z_N$ is singular. The presence of such
zeros could interfere with the convergence of the sequence $W_N$. It is the
purpose of this paper to show that in zero dimensions the delta expansion for
$W$ does in fact converge, in spite of the zeros of $Z_N$.

The specific model we consider here is the zero-dimensional analogue of the
Euclidean functional integral for a $\varphi^4$ quantum field theory, which in
this case amounts to the one-dimensional integral
$$ Z\equiv\int^{\infty}_{-\infty}{\rm d}x~{\rm e}^{-gx^4-\mu^2x^2},\eqno{(1)}$$
where $g$ is the coupling constant and $\mu$ the mass. In order to perform a
weak-coupling expansion of $Z$ in powers of $g$ it is necessary that the mass
parameter $\mu^2$ be positive; a weak-coupling expansion does not exist
otherwise. However, in the linear delta expansion the value of $\mu^2$ is
immaterial. Thus, for simplicity we restrict our attention to the massless case
$\mu=0$; the analysis of the massive case (with either sign of $\mu^2$)
 does not differ in any significant
way.

The linear delta expansion \cite{DM} has features in common with a number of
 previous
approaches \cite{CW}-\cite{SV} to improving on the convergence of ordinary
perturbation theory. It involves the introduction of an artificial parameter
$\d$ which does not appear in the original problem and which interpolates
linearly between the theory we hope to solve, with action $S$, and another
soluble theory, with action $S_0$. The interpolating action $S(\d)$ is defined
as
$$ S(\d)\equiv\l (1-\d) S_0 +\d S,\eqno{(2)}$$
so that $S(0)=S_0$ and $S(1)=S$.

Any desired quantity is evaluated as a perturbation series in powers of $\d$,
which is then set equal to 1 at the end of the calculation. At $\d=1$ the
theory
defined by $S(\d)$ is independent of the value of $\l$. However, to any finite
order $N$ in $\d$ there is a residual $\l$ dependence, and the choice of $\l$
is
in fact crucial to the convergence of the delta expansion. Indeed, if $\l$ were
taken to be a constant independent of the order $N$, the delta expansion would
have a zero radius of convergence, just as in ordinary perturbation theory.
However, it was proved in Ref.~\cite{LDEZ0} that if $\l$ is chosen as $\sqrt
{\a N}$, the sequence of approximants $Z_N(\l_N)$ converges to $Z$. The
numerical value of $\a$ is given at the beginning of Section 3. The scaling
$\l =cN^{2/3}$ was shown in Ref.~\cite{LDEZ1} to guarantee the convergence of
the corresponding sequence for the finite-temperature partition function of the
anharmonic oscillator. However, the proof did not extend to zero temperature
because the limits $\b \to \infty$ and $N\to\infty$ are not interchangeable.

In the context of field theory it is natural to work with the logarithm of the
partition function, i.e. with connected diagrams. It is therefore desirable to
extend the proof of convergence to $W=\ln Z$. To do so it is necessary to
determine the location of the zeros of $Z_N$ in the complex-$\d$ plane in the
limit of large $N$. This determination is performed asymptotically by a
steepest-descent evaluation of the integral representing $Z(\d)$ and an
asymptotic analysis of the behavior of the remainder $R_N$ as a function of
complex $\d$.

In Section 2 we explain how the zeros of $Z_N$ affect the convergence of the
$\d$ expansion for $W$. The asymptotic analysis of the location of these zeros
is given in Section 3. Finally, in Section 4 we summarize our results and
discuss possible extensions of the analysis presented here to higher
dimensions.
\newpage
\subsection*{2. Relevance of the Zeros of $Z_N$}

In this paper we are investigating the convergence of the delta
expansion for $W=\ln Z$. This involves computing the $N$th partial sum
$W_N$ of the Taylor series in $\d$ of $W(\d)=\ln Z(\d)$, where
$$ Z(\d)=\int_{-\infty}^\infty dx\;{\rm e}^{-\l x^2 +\d (\l x^2 -x^4)}~,
\eqno{(3)}$$
in which we have taken $g=1$ without loss of generality. Then $\d$ is set equal
to 1 and $\l$ chosen in some appropriate fashion as a function of $N$.

It was shown in Ref.~\cite{LDEZ0} that with the appropriate scaling of $\l$ the
sequence of approximants $Z_N(\d)$ evaluated at $\d=1$ tends to the exact
result. Consequently the sequence $\ln Z_N$ also tends to $W$. However, this
does {\sl not} constitute the systematic expansion of $W$ in powers of $\d$
that
we seek. That is to say, $W_N$, the sum of the first $N$ terms of the Taylor
expansion of $\ln Z$ is not the same as $\ln Z_N$, the logarithm of the sum of
the first $N$ terms of the Taylor expansion of $Z$.

To examine the convergence of $W_N$ to $W$ we will make use of a slight
generalization of some identities introduced in Ref.~\cite{LDEZ1}. Given a
function $F(\d)$, the  $N$th partial sum of its Taylor expansion evaluated at
$\d=1$ can be represented as the contour integral
$$ F_N(1)={1 \over 2\pi i} \oint_{C_0}{dz \over z^{N+1}} {1 \over 1-z} F(z)~,
\eqno{(4)}$$
where $C_0$ is a closed anticlockwise contour encircling the origin but not the
point $z=1$. The quantity $F(1)$ itself can be represented as
$$F(1)={1 \over 2\pi i} \oint_{C_1} \;{dz \over z^{N+1}} {1 \over z-1} F(z)~,
\eqno{(5)}$$
where $C_1$ is a closed contour encircling the point $z=1$ but not the origin.
Subtracting Eq.~(4) from (5) gives the following general integral
representation
for the remainder $R_N = F(1)-F_N(1)$:
$$R_N = {1 \over 2\pi i} \oint_{C_{01}} \;{dz \over z^{N+1}}
{1 \over z-1} F(z)~, \eqno{(6)}$$
where the contour $C_{01}$ encircles both $z=0$ and $z=1$ in an anticlockwise
direction, as shown in Fig.~1.

Now let us apply these identities to the function $F(\d)=\ln Z_N(\d)$ evaluated
at $\d=1$ in order to obtain a bound on the remainder, which in this case we
denote by
$${\cal R}_N \equiv\ln Z_N -(\ln Z_N)_N~.\eqno{(7)}$$
Expressing $Z_N(z)$ in terms of its roots:
$$Z_N(z)=Z_N(0) \prod_{r=1}^N \left ( 1-{z \over z_r} \right )\eqno{(8)}$$
so that
$$\ln Z_N(z)=\ln Z_N(0)+\sum_{r=1}^N\ln\left( 1-{z\over z_r}\right)~.
\eqno{(9)}$$
Thus,
$${\cal R}_N={1 \over 2\pi i} \sum_{r=1}^N \oint_{C_{01}} \;
{dz \over z^{N+1}} {1 \over z-1} \ln (z-z_r)~. \eqno{(10)}$$
Note that constant terms such as $\ln Z_N(0)$ do not contribute to the integral
around the contour $C_{01}$.

In the derivation of Eq. (10) we are assuming that there are no singularities
of
the integrand inside the contour other than those at $z=0$ and $z=1$. We will
verify this assumption in Section~3. Let us now expand the contour $C_{01}$ by
pushing it outward in all directions. In so doing we encounter the logarithmic
branch points emanating from the roots $z_r$. We take the branch cuts to lie
along straight lines radiating directly outwards, as shown in Fig.~2. Thus, the
shifted contour wraps around these branch cuts and the integral is then given
by
the sum of the discontinuities across each branch cut:
$$ {\cal R}_N= \sum_{r=1}^N \int_{z_r}^{\infty} \;
{dz \over z^{N+1}} {1 \over z-1}~. \eqno{(11)}$$

By parametrizing the integration variable $z$ along each branch cut by $\rho
{\rm e}^{i\theta_r}$ it is easy to establish the bound
$$|{\cal R}_N| < \sum_{r=1}^N {1\over |z_r -1|} {1\over N}
{1 \over |z_r|^{N}} ~, \eqno{(12)}$$
where we have used the inequality $|z-1| > |z_r -1|$ along each radial cut.
This latter inequality is valid provided that $|z_r| > 1$, which will be
established in the next section. If $z_{\rm min}$ denotes the root having the
smallest modulus then Eq.~(12) may be replaced by the simpler inequality
$$|{\cal R}_N|<{1\over |z_{\rm min}-1|}{1\over |z_{\rm
min}|^{N}}~.\eqno{(13)}$$

Since $|z_{\rm min}|>1$, we may conclude from Eq.~(13) that the remainder
${\cal R}_N$ vanishes for large $N$. The rate at which ${\cal R}_N$
tends to 0 is crucially dependent on the behavior of $z_{\rm min}$
as a function of $N$, which is the subject of the next section.
\newpage
\subsection*{3. Saddle-Point Determination of the Smallest Zero of $Z_N$}

In Ref.~\cite{LDEZ1} it was shown that the sequence $Z_N$ converges to $Z$ if
the parameter $\l$ is chosen as $\l =\l_N =\sqrt{\a N}$, where the numerical
value $\a =1.3254\ldots$ is obtained from $\a =2/\sinh\b$ and $\b$ satisfies
the
transcendental equation $\b =\coth\b$. We adopt the same choice for the
parameter $\l$ here. In this case the series
$$Z_N(z) = \sum_{n=0}^N c_n z^n\eqno{(14)}$$
looks rather simple: apart from the last term, which is small and negative, the
coefficients are all positive and monotonically decreasing, looking roughly
like
a geometric series. Unfortunately, it is not easy to determine the location of
the roots of a polynomial from the knowledge of the coefficients $c_n$. As an
example, consider the simple-looking case $c_n = {\rm e}^{-\g n^2}$. For small
values of $\g$ the roots lie on a circle centered at the origin in the
complex-$z$ plane. However, as $\g$ increases past a critical value two pairs
of complex conjugate roots break away from the circle. Additional roots
eventually break away from the circle as $\g$ increases through a whole
sequence
of critical values. This phenomenon is described in Ref.~\cite{ARN}. To our
knowledge there is no simple analytic way to determine these critical values
or,
indeed, the positions of the roots.

Interestingly, the configuration of the roots of $Z_N$ shares many of the
characteristics of this simple model. For large odd $N$ all but five of the
roots lie almost exactly on a circle; of the remaining roots, one
complex-conjugate pair lies inside the circle, another conjugate pair lies
outside, and there is a single root located far away on the positive-real axis
(see Fig.~3 for the case $N=27$). When $N$ is even the only qualitative
difference is that there is no real root. As shown in Eq.~(13), it is the
position of the pair of roots nearest the origin which determines whether or
not
the sequence $W_N$ converges to $W$.

As remarked above, it is very difficult to find the positions of the roots
directly from the coefficients. Thus, we have adopted the alternative strategy
of expressing $Z_N (z)$ as the difference
$$Z_N (z) = Z(z) - R_N (z) \eqno{(15)}$$
and estimating each term on the right side of (15) by asymptotic methods. Note
that while $Z(z)$ does not depend explicitly on the large parameter $N$, when
$z\neq 1$ it does involve $\l_N$, which {\sl is} a large parameter. It is
through this dependence that we are able to estimate the integral
representation
for $Z(z)$ by steepest-descent analysis. The result of this analysis is that
$$|z_{\rm min}| = 1 + \left (  {3\pi \over \a N} \right ) ^{1/2}
+ {\rm O}\left ( {1\over N}\right ) ~.\eqno{(16)}$$
{}From this asymptotic relation we conclude that ${\cal R}_N$ tends to zero
like
$\exp (-\sqrt {3\pi N/\a})$, thus proving that $W_N$ converges to $W$.

The asymptotic analysis of the two terms $R_N (z)$ and $Z(z)$ on the
right side of  Eq.~(15) follows in Subsections 3A and 3B, respectively.

\subsubsection*{A. Asymptotic Bound on $R_N(z)$}
The starting point for our analysis is the identity for $\Theta_N (y) \equiv
e^{-y} \{ e^y \}_N$ established in Ref.~\cite{LDEZ0}:
$${d\over{dy}}\Theta_N = - y^N e^{-y}/N!~.\eqno{(17)}$$
This identity is to be applied under the $x$ integration with $y=z(\l x^2 -
x^4)$. That is,
$$R_N(z) = 2\int_0^{\infty}dx\; e^{-\l x^2 +y} [1-\Theta_N (y)]~.\eqno{(18)}$$

Let us first deal with the case ${\rm Re}~z>0$, where the $x$ integration
contour can be maintained along the real axis. For complex $z$ in this region
we
integrate Eq.~(17) along the ray $y' = yt$, with $0\leq t\leq 1$, to obtain
$$\Theta_N (y) = 1 - \int_0^y dy'\; y'^N e^{-y'}/N!~,\eqno{(19)}$$
using $\Theta_N (0) = 1$.

The $x$ integration in Eq.~(18) splits naturally into the two ranges, $0\leq x
\leq\sqrt{\l}$ and $\sqrt{\l} \leq x$. In the first case $y = |y| e^{i\theta}$,
where $\theta={\rm arg}(z)$. The contribution to the remainder from this range
is therefore
$$A_N(z)= {2\over{N!}}\int_0^{\sqrt{\l}}dx\;e^{-\l x^2+y}\int_0^{|y|}d\omega
\;\omega^N e^{-\omega e^{i\theta}} e^{i(N+1)\theta}~.\eqno{(20)}$$
Thus,
$$|A_N(z)|\leq {2\over{N!}}\int_0^{\sqrt{\l}}dx\;e^{-\l x^2+|y|\cos\theta}
\int_0^{|y|}d\omega \;\omega^N e^{-\omega \cos\theta}~.\eqno{(21)}$$

The maximum of the $\omega$ integrand occurs at $\omega = N/\cos\theta \geq N$,
whereas the upper limit is less than or equal to $N\alpha |z|/4$. Thus,
provided that $|z| < 4/\alpha$, we may bound $|A_N(z)|$ by
$$|A_N(z)|\leq {2\over{N!}}\int_0^{\sqrt{\l}}dx\;e^{-\l x^2+|y|\cos\theta}
|y|^{N+1} e^{-y\cos\theta}$$
$$\quad\quad\quad\quad\quad
={2\over{N!}} |z|^{N+1} \int_0^{\sqrt{\l}}dx\;e^{-\l x^2}
(\l x^2 - x^4)^{N+1}~.\eqno{(22)}$$

Apart from the factor of $|z|^{N+1}$, the right side of Eq.~(22) is
precisely $A_N (1)$, which was shown in Ref.~\cite{LDEZ0} to be bounded by
$CN^{3/4}e^{-N\a/2}$. Thus,
$$|A_N(z)| \leq CN^{3/4}|z|^{N+1} e^{-N\a/2}~.\eqno{(23)}$$

In the second of the two regions of the $x$ integration, the quantity $\l x^2-
x^4$ is negative, so that $y = \omega e^{i(\theta + \pi)}$. The
corresponding contribution to the remainder is then
$$B_N(z) = {2\over{N!}}\int_{\sqrt{\l}}^{\infty}dx\;e^{-\l x^2+y} \int_0^{|y|}
d\omega \;\omega^N e^{\omega e^{i\theta}} e^{i(N+1)(\theta
+\pi)}~,\eqno{(24)}$$
so that
$$|B_N(z)|\leq{2\over{N!}}\int_{\sqrt{\l}}^{\infty}dx\;e^{-\l
x^2-|y|\cos\theta}
\int_0^{|y|} d\omega \;\omega^N e^{\omega \cos\theta}~.\eqno{(25)}$$
In this case the $\omega$ integrand is a monotone function whose maximum
occurs at the upper limit. Thus,
$$|B_N(z)|\leq{2\over{N!}}\int_{\sqrt{\l}}^{\infty}dx\;e^{-\l
x^2-|y|\cos\theta}
|y|^{N+1} e^{|y| \cos\theta}$$
$$={2\over{N!}}|z|^{N+1} \int_{\sqrt{\l}}^{\infty}dx\;e^{-\l x^2}
(x^4 - \l x^2)^{N+1}~.\eqno{(26)}$$
Again, the right side is $B_N(1)$ multiplied by a factor of $|z|^{N+1}$. In
Ref.~\cite{LDEZ0}, $B_N(1)$ was shown to be bounded for large $N$ in precisely
the same way as $A_N(1)$.

Thus, altogether
$$|R_N(z)| \leq CN^{3/4}|z|^{N+1} e^{-N\a/2}~,\eqno{(27)}$$
provided that
$$|z| \leq {4\over \a}~.\eqno{(28)}$$

Now let us consider ${\rm Re}\;z <0$. The only difference is that since
$\cos\theta < 0$ the roles of $A_N$ and $B_N$ are reversed. That is,
$|A_N(z)|$ is bounded as in Eq.~(23) independent of $|z|$, while $|B_N(z)|$
is bounded in the same way, provided that Eq.~(28) holds. Altogether, the
conclusion for $|R_N(z)|$ remains the same.

\subsubsection*{B. Steepest-Descent Evaluation of $Z(z)$}

For ${\rm Re}\;z > 0$ the integral in Eq.~(3) is an adequate definition of the
function $Z(z)$ in the complex-$z$ plane. Later on in this subsection we will
examine the analytic continuation of $Z(z)$ to the left-half $z$ plane.
Because $\l=\l_N$ is large we evaluate the integral by looking for the saddle
points of the exponent $\varphi$ in the integrand:
$$\vf (x) \equiv -\l x^2 + z (\l x^2 -x^4)~.\eqno{(28)}$$
The three saddle points satisfying $\vf'(x) = 0$ are $x_0=0$ and
$x_{\pm}= \pm \sqrt {\l (z-1)/(2z) }$. At these saddle points
$\vf (x_0) = 0$ and $\vf (x_{\pm}) = -\l^2 (z - 1)^2/(4z)$.

The method of steepest descents requires that for each value of complex $z$ we
deform the integration path in Eq.~(3) to a stationary-phase contour in the
complex-$x$ plane connecting the original end points at $x=\pm\infty$. We find
that there are two cases to consider. When $|z|<1$ (region \cA) the
stationary-phase contour passes through the saddle point $x_0$ only and not
through the others. In contrast, when $|z|>1$ the stationary-phase contour
passes
through all three saddle points. The appropriate integration contours for these
two cases are illustrated in Figs.~4 and 5, respectively. In the second case
there are two subregions in the complex-$z$ plane to consider, region \cB, in
which $\exp [\vf (x_\pm )]$ is subdominant with respect to $\exp [\vf
(x_0)]=1$,
and region \cC~ in which the reverse is true. The boundary curve $\Gamma$
between subregions \cB~ and \cC~ is given by the polar equation
$$\Gamma:\quad\cos \th = {2r \over {r^2 + 1}}~, \eqno{(29)}$$
where $z =r{\rm e}^{i\th}$. The above regions and the boundary curves are
illustrated in Fig.~6.

In regions \cA~ and \cB~ the dominant contribution to $Z_N(z)$ comes from the
saddle point at the origin, whose contribution is a slowly-varying
(nonexponential) function of $N$: $Z(z) \sim N^{-1/4}$. Referring to Eq.~(15),
we see that a zero of $Z_N(z)$ must arise from a cancellation between this
contribution and $R_N(z)$, which as we have seen is bounded by ${\rm const}
\times~
\exp [N(\ln |z| - \a/2)]$. Thus, no zero is possible for $|z| < {\rm e}^{\a/2}
=1.94\dots$. Since this value is strictly greater than one, the zeros in
regions
\cA~ and \cB~ will not affect the convergence of ${\cal R}_N$. From the
numerical plot in Fig.~3 it appears that the bound on $|z|$ is saturated,
giving rise to a ring of zeros.

In region \cC~ the dominant contribution to $Z_N(z)$ comes from the saddle
points $x_\pm$. These saddle points give a nontrivial exponential dependence to
$Z_N(z)$ of the form $\exp [N\a (z-1)^2/(4z)]$. Since this exponent is
positive the necessary cancellation required for a zero can only occur at still
larger values of $|z|$. This accounts for the conjugate pair of complex zeros
lying outside the ring in Fig.~3 and the large real root. None of these zeros
affects the convergence of ${\cal R}_N$.

Finally, we examine the boundary curve $\Gamma$ separating regions \cB~ and
\cC.
On this curve the exponent of the saddle points at $x_\pm$ is the same as that
of $x_0$. There thus arises the possibility of a cancellation between their
contributions, leaving a remainder which can be compensated by $R_N(z)$ at a
smaller value of $|z|$ inside the ring of zeros in Fig.~3. Note that the
smallest zero marked in Fig.~6 lies almost exactly on $\Gamma$, indicating that
this is indeed the mechanism that produces the zero of smallest absolute value.
In order for this cancellation to occur the two saddle-point contributions must
be completely out of phase. The phase of each contribution is of course
uniquely
determined by the direction of the stationary-phase path going through the
saddle point. As can be seen from Fig.~5, the stationary-phase path directions
differ by $\pi/2$. Thus, the required condition for cancellation is
$${\rm Im}\;\vf (x_\pm) = \pi/2 + (2n+1)\pi~,\eqno{(30)}$$
where $n$ is any integer.

This equation is to be solved in conjunction with Eq.~(29). The condition $|z|>
1$ excludes negative values of $n$ and in fact the smallest value of the
modulus
of the root $z$ is obtained for $n=0$. With $n=0$ the simultaneous solution of
Eqs.~(29) and (30) is
$$z = a + \sqrt{a^2 - 1}~, \eqno{(31)}$$
where $a=1+3\pi i/(N\a)$. The accuracy of our asymptotic analysis improves with
increasing $N$. For $N=27$ this saddle-point analysis predicts the positions of
the smallest roots quite accurately: $z =  1.481\ldots \pm (0.811 \dots)i$, to
be compared with the actual numerical values $z =1.443\ldots\pm (0.835\dots)i$,
as shown in Fig.~6. For larger values of $N$ the roots approach their
asymptotic
values with an error that behaves like $1/\sqrt{N}$.

{}From Eq.~(31) we can determine the behavior for large $N$ of the modulus of
the
smallest root $z_{\rm min}$. This is given in Eq.~(16), which we have checked
numerically by performing a fit to a series in inverse powers of $\sqrt{N}$ up
to $N=59$. The crucial feature of Eq.~(16) is that $|z_{\rm min}|$ approaches 1
from above sufficiently slowly that ${\cal R}_N$, the difference between $W_N$
and $W$, tends to zero. While the convergence is not as rapid as that of
the sequence $Z_N$, which converges like $\exp (-\a N/2)$, it still converges
like an exponential: $\exp (-\sqrt { 3 \pi N/\a} )$.

Up till now we have not considered the case $\Re\; z < 0$. For this case the
integral representation for $Z(z)$ is no longer valid as it stands because it
is divergent. As $z$ is rotated into the left-half complex plane the endpoints
of the integration contour must be rotated in the opposite direction (and at
one quarter of the rate) in order for the integral continue to to exist. A
description of this analytic continuation procedure may be found in
Ref.~\cite{BT} for the case of boundary conditions on differential equation
eigenvalue problems. Once the integral representation for $Z(z)$ has been
continued to negative values of $\Re\; z$ it may be subjected to the same
saddle-point analysis as above. For this case, there are just two regions,
$|z|<1$ in which the stationary-phase contour passes only through the saddle
point $x_0$ and $|z| > 1$ in which the contour passes through all three saddle
points, but the contribution from $x_0$ dominates. These two cases correspond
to
what happens in regions \cA~ and \cB~ for $\Re\; z >0$. Thus, the resulting
zeros
complete the ring of zeros on Fig.~3 but do not affect our conclusions
regarding
the convergence of $W_N$.

\newpage
\subsection*{4. Discussion and Conclusions}

The arguments presented above establish the convergence and bound the remainder
in an optimized expansion of $W\equiv\ln{Z}$ for the non-Gaussian integral (3),
provided we choose to vary the optimizing parameter $\l$ with $N$ at large $N$
so as to optimize the convergence of the partials $Z_N$ (i.e. we take $\l\sim
\sqrt{N}$):
$$\ln{Z}-\ln{Z_N} \simeq e^{-N\a/2}$$
(ignoring power prefactors). With this choice of $N$ dependence for $\l$, we
have shown that
$$\ln{Z_N} -(\ln{Z_N})_{N} \simeq e^{-\sqrt{3\pi N/\a}}~.$$
The remainders in the partials $W_N\equiv (\ln{Z_N})_{N}$ for the connected
function $W$ are thus asymptotically
$$W-W_N \simeq e^{-\sqrt{3\pi N/\a}}~.$$

Although the numerical illustrations have been for odd $N$, we should emphasize
that with $\l$ chosen as $\sqrt {\a N}$ the results for even $N$ interpolate
smoothly between those for odd $N$. The distinction originally arose
\cite{LDEZ1}
in the context of the principle of minimal sensitivity (PMS, \cite{PMS}) as
applied to $Z_N(\l)$. There it turns out that for odd $N$ there is a single
stationary point of $Z_N$ as a function of $\l$, whereas for even $N$ there is
a
point of inflection but no stationary point. The present paper marks a step
away
from reliance on the PMS philosophy. Instead we are simply choosing $\l$ in
such
a way that the convergence of the sequence $W_N$ is guaranteed. This
methodology
eliminates the arbitrariness that often occurs in applying the PMS criterion.
That is, there may well be several stationary points \cite{FC}, and one then
has
to choose between them by some further criterion which itself needs to be
justified. In the case of $W_N$ there are indeed several stationary points, and
one no longer has a strict inequality like $Z_N < Z$ to help one distinguish
between them. For example, for $N=19$ there are two maxima and one minimum in
$\l$. The first maximum and minimum, illustrated in Fig.~7, are reasonably
close to the exact value of $W$, while the second maximum exceeds it by
some $0.4\%$. The value of $\l$ given by $\l=\sqrt{\a N}$ is slightly lower
than the position of the first maximum, and gives a better estimate of
$W$: 0.5948757 compared with the exact value of $0.5948753\ldots$.

The convergence of the optimized expansion for $W$ is slower than that for $Z$
with this choice of $\l(N)$. It is of course possible that a more rapid
convergence (with $\ln(W-W_N) \simeq -N^{\nu}$, $\frac{1}{2} <\nu<1$) might be
obtained with a different choice of $\l(N)$. Our main object here has been to
provide an existence proof for a convergent procedure for the connected
generating function. An understanding of the convergence at the level of
connected quantities is crucial in higher dimensions, where the delta expansion
for the full partition function converges at any finite spacetime volume, but
at
a rate which deteriorates as the volume is increased. As $W$ is linear in the
volume (for large volume) a convergent procedure at any finite volume will be
uniformly convergent (for connected quantities) as the volume cutoff is
removed.
Finally, we note that the techniques used above involve only saddle-point
estimates which should generalize readily to functional integrals defining the
partition functions for quantum mechanics or field theories.

\newpage
\subsection*{Figure Captions}
\begin{description}
\item[Fig.~1] Contours in the complex-$z$ plane used for the representation
of $F_N(1)$, $F(1)$, and $R_N$ (Eqs.~(4)-(6)).
\item[Fig.~2] Logarithmic branch cuts radiating from the zeros of $Z_N(z)$
in the representation of ${\cal R}_N$ in Eq.~(11).
\item[Fig.~3] Location of zeros of $Z_N(z)$ in the complex plane for $N=27$.
The circle (of radius 2.1) is just an empirical fit to the ring of zeros.
The distant real root at $z =  14.124$ is not shown.
\item[Fig.~4] Stationary-phase contours of the function $\vf(x)$ in Eq.~(28)
for a typical complex value of $z$ with $|z| < 1$. The original integration
contour lying along the real axis must be distorted into the contour
marked $Ax_0B$, passing through the saddle point $x_0$. The other saddle points
play no role in the asymptotic evaluation of the integral.
\item[Fig.~5] Stationary-phase contours of the function $\vf(x)$ in Eq.~(28)
for a typical complex value of $z$ with $|z|>1$. The original integration
contour lying along the real axis must be distorted into the contour marked
$Ax_-Bx_0Cx_+D$, passing through the three saddle points $x_0$ and $x_\pm$.
\item[Fig.~6] Right-half $z$ plane showing the three regions \cA, \cB, \cC~
discussed in Subsection 3B. The boundary between regions \cB~ and \cC, labeled
$\Gamma$, is given in Eq.~(29). Also shown are the two smallest-modulus
roots of $Z_N(z)$ for $N=27$.
\item[Fig.~7] $W_N(\l)$ versus $\l$ for $N=19$, showing the lower maximum and
the minimum. The dotted line is at $\l = \l_N = \sqrt{\a N}$. Note that the
vertical scale is highly magnified, with a range of about $7\times 10^{-5}$.
\end{description}
\end{document}